\documentclass[manuscript]{aastex}

\newcommand{\kms}{{\>\rm km\>s^{-1}}}

\newcommand{\sprocess}{{\it s}-process}

\begin{document}

\title{William Pendry Bidelman (1918--2011)\altaffilmark{1}}

\author{Howard E. Bond\altaffilmark{2}}

\altaffiltext{1}{Material for this article was contributed by several family
members, colleagues, and former students, including: Billie Bidelman Little,
Joseph Little, James Caplinger, D. Jack MacConnell, Wayne Osborn, George W.
Preston, Nancy G. Roman, and Nolan Walborn. Any opinions stated are those of the
author.}

\altaffiltext{2}{Department of Astronomy \& Astrophysics, Pennsylvania State
University, University Park, PA 16802; heb11@psu.edu}

\begin{abstract}

William P. Bidelman---Editor of these {\it Publications\/} from 1956 to
1961---passed away on 2011 May 3, at the age of 92. He was one of the last of
the masters of visual stellar spectral classification and the identification of
peculiar stars. I review his contributions to these subjects, including the
discoveries of barium stars, hydrogen-deficient stars, high-galactic-latitude
supergiants, stars with anomalous carbon content, and exotic chemical abundances
in peculiar A and B stars. Bidelman was legendary for his encyclopedic knowledge
of the stellar literature. He had a profound and inspirational influence on many
colleagues and students. Some of the bizarre stellar phenomena he discovered
remain unexplained to the present day.

\end{abstract}


\clearpage

William Pendry Bidelman---famous among his astronomical colleagues and students
for his encyclopedic knowledge of stellar spectra and their
peculiarities---passed away at the age of 92 on 2011 May 3, in Murfreesboro,
Tennessee. He was Editor of these {\it Publications\/} from 1956 to 1961. 
Bidelman was born in Los Angeles on 1918 September~25, but when the family fell
onto hard financial times, his mother moved with him to Grand Forks, North
Dakota in 1922. There he was raised by his grandparents.\footnote{Bidelman's
maternal grandfather was Joseph Bell DeRemer, a noted architect who designed a
number of important buildings in North Dakota, including the state capitol.
Although his father had an identical name, Bidelman did not append a ``Jr.''\ to
his own name after his college days. His mother was a pianist who traveled with
dancer Martha Graham and the Denishawn troupe, described at the time as
America's leading dance company.} At age 10, he became smitten with classmate
Verna P. Shirk, later to become his wife of 69 years. Bidelman developed an
interest in astronomy at an early age; he reminisced many decades later that as
a child he had written to Alfred H. Joy at the Mount Wilson Observatory to ask
how he could become an astronomer, and had received an encouraging reply.

He attended Harvard College, graduating in the Class of 1940 (with classmate
John F. Kennedy). The day before graduation, he and Verna were married in the
Harvard Chapel (Figure~1). Bidelman then enrolled in graduate school at the
Yerkes Observatory, University of Chicago. There he was trained in stellar
spectroscopy by giants of the subject, including O.~Struve, P.~C. Keenan, and
his thesis advisor, W.~W. Morgan. His Ph.D. dissertation, a spectroscopic study
of the Double Cluster in Perseus, appeared in {\it The Astrophysical Journal\/}
(Bidelman 1943). During the war years, he served as a physicist at the U.S.
Army's Ballistic Research Laboratory at the Aberdeen Proving Ground in Maryland.
He returned to Yerkes after World War~II as an assistant professor (Figure~2).
Later in his career, Bidelman taught and conducted research at the University of
California's Lick Observatory, the University of Michigan, the University of
Texas, and lastly at Case Western Reserve University.

One of his first post-war papers (Morgan \& Bidelman 1946), which pointed out
that the A-type stars in the North Polar Sequence have very small interstellar
reddening, attracted little attention then, or in subsequent years. Yet many
years later, Morgan (1978) recalled that this had been ``the principal paper
along the way toward the {\it UBV\/} system.''  In view of the seminal
importance of {\it UBV\/} photometry in the ensuing decades, this short paper
has a remarkably high ratio of scientific impact to number of citations.

Universally known as ``Billy," Bidelman was one of the last of the great masters
of stellar spectral classification and the recognition of spectroscopic
anomalies. Younger astronomers of today may be only dimly aware of the situation
60 or 75 years ago, when stellar spectra presented a bewildering array of poorly
understood and sometimes outright bizarre phenomena, often with little or no
physical explanation. The morphological approach to spectral classification,
into which pioneers like A. Maury, A.~J. Cannon, C.~Payne, H.~N. Russell, and
Morgan had such insight, was the key to progress. First a network of ``normal''
stars has to be established. Only then can those unusual objects that depart
from this pattern of normality---those that most ruthlessly expose our ignorance
of stellar astrophysics---be recognized and studied.

An early example of this approach is Billy's discovery, along with his colleague
Keenan, of the class of ``barium'' stars (Bidelman \& Keenan 1951). These are
red giants that do not fit into the sequence of stars with normal spectra. They
exhibit a strong absorption line of \ion{Ba}{2} 4554 \AA, along with stronger
features of \ion{Sr}{2}, CH, CN, and the C$_2$ molecule than in spectra of
normal red giants. At the time, it was an utter mystery why the chemical element
barium would have an apparently enhanced abundance in certain rare stars, and
what this might have to do with a high content of strontium and carbon. The
developing science of nuclear astrophysics provided a partial explanation later
in the 1950's, in terms of the ``\sprocess" of neutron-capture reactions,
occurring in the interiors of highly evolved stars, which synthesizes heavy
elements like Sr and Ba. But it was already known to Bidelman and Keenan that
barium stars are relatively normal giants, not the highly evolved luminous ones
that the theorists said were required. It took the eventual discovery that
essentially all barium stars are spectroscopic binaries with relatively long
orbital periods (e.g., McClure 1984) to provide the modern explanation. A barium
star is the companion of a primary component that did become a highly evolved,
\sprocess-enhanced object, whose stellar wind transferred this altered material
to the surface of the barium star. The former primary is now an optically
invisible white dwarf.

In the same year, Billy published another important paper (Bidelman 1951),
pointing out the existence of stars of spectral types A and F at high Galactic
latitudes (such as 89~Herculis and HD~161796) whose spectra closely resemble
those of massive, high-luminosity supergiants. Luminous supergiants are normally
found only in the Galactic plane. This anomaly eventually led to the recognition
that high-latitude ``supergiants'' are actually lower-mass and less luminous
stars evolving off of the asymptotic giant branch (post-AGB stars). In the same
paper, Bidelman pointed out the remarkable spectrum of HR\,885, a red giant
displaying very weak features of CH and CN\null. To my knowledge, a completely
convincing explanation of the rare carbon-deficient giants like HR\,885 does not
yet exist (e.g., Palacios et al.\ 2016).

As a staff member at Yerkes, Bidelman spent long spectroscopic observing runs at
McDonald Observatory in west Texas---a location even more isolated then than it
is now. One result of these studies was his recognition (Bidelman 1952, 1953)
that there are stars across a wide range of spectral types that are
spectacularly deficient in hydrogen. In the spectrum of the O-type star HD
160641, Billy's low-dispersion spectrograms showed no trace of the Balmer lines,
but there were strong features of helium and of doubly ionized carbon and
nitrogen. Bidelman's star is now classified an an ``extreme helium star"---one
of the hottest ones known, according to Wright et al.\ (2006). Current
explanations (e.g., Saio \& Jeffery 2002) of such stars involve either a
post-AGB star that has lost its entire hydrogen envelope, probably during a late
helium core flash; or that they have an origin in mergers of binaries containing
a helium white dwarf and a carbon-oxygen white dwarf. (It is worth noting that
the preceding sentence would have been almost entirely incomprehensible in
1953---such is the progress of astrophysical theory, building as it does on
observational challenges.)

Bidelman's familiarity with the literature of stellar spectroscopy was
legendary. Nancy Roman remembers that ``Billy's fantastic memory played a role
in my career.'' She had found remarkable variations in the spectrum of the star
now known as AG~Draconis, and mentioned them to Bidelman. He remembered what the
spectrum looked like when he had observed it some years earlier, which made it
possible for the changes to be described in some detail (Roman 1953). It was
this contribution, she recalls, that brought her to the attention of scientists
at the Naval Research Laboratory, who later became the core of the scientific
staff at NASA\null. Subsequently they asked Nancy to set up a program in space
astronomy. As has been told elsewhere\footnote{e.g., \tt
https://www.astrosociety.org/wp-content\slash uploads/2013/07/ab2013-112.pdf},
Nancy Roman became a key leader in NASA astrophysics, up to and including the
spectacular success of the {\it Hubble Space Telescope.}

I myself recall, during a discussion of a particular high-velocity star that we
had in the department library at Michigan, that he went right to the shelf full
of ApJ's, pulled down a volume, and opened it to the very page that gave the
information that we needed on that star---no need to waste time searching for
that paper in the five-year index! On another occasion, I had come across an
M~supergiant with strong [\ion{Fe}{2}] emission and other peculiarities while
visually scanning an objective-prism plate. After some effort in the library, I
was able to identify the object with a known variable star, WY~Velorum. I then
proudly took the plate to Billy for his reaction to my great discovery, without
mentioning that I already knew its identity. His reaction: ``This is amazing!
Congratulations! I only know of one other star in the sky with a spectrum like
this---WY~Velorum!'' 

The story told among the graduate students in those days was that, if you were
asked on your preliminary examination how you would go about determining the
spectral type of a given star, the correct answer was ``write down its HD
number, and go ask Bidelman.'' His deep knowledge of the literature was on
display in two review articles, ``On the Carbon and S-Type Stars'' (Bidelman
1954a) and ``The Carbon Stars---An Astrophysical Enigma'' (Bidelman 1956). Some
of the enigmas he described six decades ago remain unsolved---for example, the
still-unexplained origin of the R-type carbon stars (e.g., Dom{\'{\i}}nguez et
al.\ 2010). Speaking many years later of the first of these articles, Robert
Wing (2000) remarked ``Papers that have great impact on individuals are not
always the well-known classics. $\dots$ I have seldom seen citations to that
paper, but its effect on me was profound when I stumbled across it as a graduate
student in 1964. $\dots$ That paper influenced the direction of my dissertation,
and consequently of my entire career.''

For many decades, Bidelman maintained a card catalog of information on
individual stars, with the information and literature references entered on
$3\times5$-inch index cards. An early result was his ``Catalogue and
Bibliography of Emission-Line Stars of Types Later than B'' (Bidelman 1954b),
containing information on nearly 1200 stars and 903 literature references---and
published in volume~1 of {\it The Astrophysical Journal Supplement Series}. Six
decades later, this compilation still contains useful information, even in these
days of online catalogs. Billy would always respond to requests for data from
his card catalog, but it was felt desirable to make this wealth of information
more widely available. During the 1970s, the contents of over 60,000 index cards
from Bidelman's collection were converted to machine-readable data by manually
keypunching  (Parsons et al.\ 1980, 1996). These data are now part of the online
data available to all astronomers at the Strasbourg VizieR site\footnote{\tt
http://vizier.u-strasbg.fr/viz-bin/VizieR?-source=VI/32}. 

Bidelman moved to Lick Observatory in 1953, accompanied by Verna and their four
daughters (Figures~3 and 4). They were assigned a large rambling house (as
described by his daughter Billie Jean) on Mount Hamilton---from which the Bay
Bridge could be seen from the living-room windows.  The girls had the run of the
mountain, where they attended a one-room school, with about a dozen other
children of all ages. One spring, when Billie was about six, she ran home with a
bouquet of 20 wild yellow-orange California poppies picked from around the
120-inch dome. Her father informed her that she owed the government \$1,000, as
the fine for picking the state flower was \$50 apiece. The Bidelmans threw a New
Year's Eve party every year---from which fortunately everyone could walk home
when the party finally broke up. Many visiting astronomers were entertained in
the Bidelman household, sometimes at the last minute---to the consternation of
Verna, since the nearest grocery store was an hour's drive away.

George Preston recalls that, at the end of his first year of graduate school, he
spent the summer in residence at Lick. One night at the 36-inch refractor,
Preston accompanied Billy to the darkroom to develop spectrograms, where he
``witnessed an extraordinary event that must have affected my career.'' On
examining a plate fresh out of the hypo, Bidelman remarked that the star was a
high-velocity giant. Preston was astonished---how could someone know a star had
a high velocity when there wasn't even a comparison spectrum on the plate? ``Of
course Billy was making the connection between weak lines and high velocity, of
which I was unaware in 1955.'' A subsequent higher-dispersion spectrogram of the
metal-poor star in question---the now well-known HDE~232078---revealed that it
indeed has a very high radial velocity of about $-390\kms$ (Preston \& Bidelman
1956). Preston says that this experience aroused his interest in the newly
recognized halo population of the Milky Way and stimulated a number of his
subsequent investigations.\footnote{As a young graduate student myself, I and
several others accompanied Bidelman---an inveterate punster---to a night of
observing at Michigan's Schmidt telescope. In the darkroom, Billy poured out the
developer, and then asked ``Where's Arthur?'' We students looked at each other.
None of us were named Arthur. Then Billy said, ``Ah, here's our thermometer.''} 

Beginning around 1960, at about the time the coud\'e spectrograph came into
operation at the Lick 120-inch, Bidelman was the leading discoverer of
extraordinary abundances of exotic chemical elements in the photospheres of
peculiar A and B stars. One of the first was the bizarre spectrum of
3~Centauri~A (Bidelman 1960a), exhibiting strong absorption lines of ionized
phosphorus. Billy made this finding in the course of a systematic inspection of
photographic spectra accumulated over many decades at the Lick Observatory and
its Chile Station. The first spectroscopic plate of 3~Cen~A had been obtained
around 1910; but recognition of its peculiarities had to await an astronomer
with a broad enough knowledge of stellar spectra to identify its striking
departure from normal. More discoveries of freakish chemical abundances in Ap
stars followed quickly.  Strong lines of \ion{Ga}{2} were found in 3~Cen~A and
$\kappa$~Cancri (Bidelman 1960b), the first time gallium had been seen in
stellar spectra, although it took precise new laboratory wavelength measurements
at the National Bureau of Standards to verify this identification (Bidelman \&
Corliss 1962). Both stars also show abnormally strong lines of \ion{Mn}{2}.
Another bizarre magnetic Ap star, HR\,465, was found on a Lick spectrogram to
exhibit extraordinarily strong features of heavy elements such as molybdenum,
niobium, and neodymium (Bidelman 1962).  Lawrence Aller told me that, in those
days, colleagues would ask each other after every monthly bright-of-the-moon
spectroscopic time ``What has Bidelman discovered now?''

Many of his discoveries never appeared in the archival literature, but were
presented at meetings or communicated privately to colleagues with access to
large telescopes with coud\'e spectrographs. Among the most baffling findings
that resulted were the outlandishly high abundances of $^3$He, krypton, and
xenon in 3~Cen~A (e.g., Sargent \& Jugaku 1961; Jugaku et al.\ 1961). A short
conference presentation (Bidelman 1966a) gave some results from his line
identifications in high-dispersion Lick spectrograms of Ap stars, including
anomalously high abundances of selenium, palladium, praseodymium, and perhaps
most astonishingly, mercury. The effect is not subtle: the abundance of Hg can
be 400,000 times the solar value. More amazingly, the mixture of isotopes of
mercury (which can be determined from the isotopic hyperfine splitting of the
\ion{Hg}{2} 3984~\AA\ line, using very high-dispersion spectrograms of very
slowly rotating stars) varies from star to star (e.g., White et al.\ 1976). In
unpublished work cited by Fuhrmann (1989), Bidelman had identified osmium,
platinum, gold, and bismuth, among other extremely heavy elements, in the
spectrum of HR\,465---a magnetic Ap star with a rotation period longer than two
decades.

These stunning chemical abundances were, to astronomers of the mid-1960's, a
profound astrophysical mystery comparable to the present-day problem of dark
energy. The anomalies seemed to expose a woeful ignorance of the origin of a
basic stellar phenomenon. The first international conference on peculiar A
stars, held at the Goddard Space Flight Center in 1965, attracted the leading
stellar astronomers of that era. Billy was invited to give the introductory
overview (Bidelman 1967). He led off with remarks that illustrate his sardonic
sense of humor (along with his knowledge of classical music) so nicely that I
quote them here:

{\narrower 

``After all, in some ways, an introduction is a little bit like an overture to
an opera. For instance, the overture isn't expected to be taken very seriously,
because people are just supposed to be sliding---mentally or physically---into
their seats. Furthermore, all the overture does is to give you bits and
snatches of the tunes that are to be played, with ample elaboration, later on
during the opera.

``At first I felt that this meant that I really shouldn't say very much. Then it
occurred to me, following the analogy a bit further, that it often happens that
the overture is the only part of the opera that survives.''

}

\noindent As Miroslav Plavec (1976) later remarked, Bidelman's ``introduction
then was really a masterpiece which can be studied independently and with great
profit.''

Theoreticians of the day struggled to explain these ``chemically peculiar''
stars. A widely cited paper (Fowler et al.\ 1965) proposed a complex scenario
that placed the Ap stars in an advanced evolutionary stage (or else they were
contaminated companions of such stars) in which an $r$-process of
nucleosynthesis and surface spallation were invoked. Unfortunately, it was known
even then that Ap stars are most likely on or near the main sequence; that they
do not have a high incidence of binaries; and that they  exhibit no evidence for
surface flaring activity. Nucleosynthetic scenarios mostly fell by the wayside
when Michaud (1970) argued that diffusive processes are primarily responsible
for the peculiar abundances in Ap stars. In a very stable atmosphere, especially
if the stability is enhanced by a strong magnetic field---as observed in many Ap
stars---a mixture of gravitational settling and radiative levitation can lead to
very large under- and overabundances of individual species in the stellar
photospheres. This explanation is now widely accepted, and peculiar~A stars
receive much less attention than they did 50 years ago. However, it remains
unclear whether diffusive processes can account for all of their oddities, such
as the wide range in isotopic abundances of mercury (see Woolf \& Lambert 1999).
Billy himself, in his final words on the subject (Bidelman 2002, 2005), had come
to believe that the Ap stars had a special origin after all (perhaps merged
binaries). He had a strong suspicion of the presence of radioactively unstable
elements with very short half-lives in the atmospheres of objects like 
HD\,101065, the infamous Przybylski's Star (Cowley et al.\ 2004). If so, these
elements must have been created very recently.

Bidelman moved east to join the University of Michigan astronomy faculty in
1962. A 24/36-inch Schmidt telescope, named in honor of Michigan astronomer H.
D. Curtis, had been installed at a relatively dark site northwest of Ann Arbor
in 1950. The Curtis Schmidt was equipped with a pair of prisms
($4^\circ+6^\circ$), providing a nearly uniquely high dispersion
($108\,$\AA$\rm\,mm^{-1}$ at H$\gamma$) for objective-prism spectroscopic
surveys (Miller 1953). In good seeing, the spectral resolution at H$\delta$ is
about 2~\AA, similar to that used for MK classification. However, apart from a
study of CN strengths in red giants by Yoss (1961), based on a collection of
several hundred objective-prism plates in four declination zones, this powerful
capability had remained largely unexploited at the time of Billy's arrival at
Michigan. Using the plate archive, he showed (Bidelman 1966b) that accurate
spectral classification was possible using such material, along with the
recognition of peculiar A and B stars, and metal-deficient late-type stars. At
his direction, several Michigan students used archival and newly obtained Curtis
Schmidt plates in their dissertation work (e.g., Bond 1970; Schmitt 1971). But,
thinking on a grander scale, Billy began to advocate a new all-sky
objective-prism survey that would reclassify all of the stars of the {\it Henry
Draper Catalogue\/} onto the MK system. 

To facilitate this goal, Bidelman was instrumental in arranging for the Curtis
Schmidt to be relocated in 1967 to Cerro Tololo Inter-American Observatory, and
a systematic survey of the entire southern sky with the $10^\circ$ prism
combination was begun. The monumental task of actual visual classification of
the spectra was taken up by Nancy Houk. The first catalog of two-dimensional MK
spectral types for the HD stars south of $-53^\circ$ appeared in 1975 (Houk \&
Cowley 1975) The southern-hemisphere survey was eventually completed to
$\delta=+5^\circ$ (Houk \& Swift 1999), with a total of more than 160,000 MK
classifications---the largest number, by far, ever carried out by one
person\footnote{Many of the Curtis Schmidt southern-hemisphere survey plates are
now archived at the Pisgah Astronomical Research Institute (Osborn \& Robbins
2009). Appallingly, most of the plates obtained in Michigan before 1967 were
discarded (W. Osborn, private communication).}.

As the objective-prism plates arrived in Ann Arbor, they were searched for
interesting objects by Bidelman, Jack MacConnell, and several graduate
assistants, including myself (see Bidelman \& MacConnell 1968; Bidelman et al.\
1973). Discoveries of many hundreds of peculiar stars of various types in this
``early-result'' program were published by Bidelman \& MacConnell (1973). Later
nearly 150 new luminous O stars and later-type supergiants were discovered in
this material by MacConnell \& Bidelman (1976), and over 175 additional peculiar
stars (Bidelman 1981; Bidelman \& MacConnell 1982).

In 1969, Bidelman took a faculty position at the University of Texas, and then
in 1970 became the Director of the Warner and Swasey Observatory at Case Western
Reserve University, where he spent the rest of his career.  Warner and Swasey's
Burrell Schmidt telescope is a twin of the Curtis Schmidt. In 1979, the Burrell
Schmidt was relocated to Kitt Peak National Observatory, and the telescope was
equipped with a new (single) $10^\circ$ objective prism for the purpose of
extending the southern objective-prism survey to the other half of the sky (Houk
\& Bidelman 1979). The survey of the northern hemisphere began in 1981, and the
observations were completed in 1992. Bidelman began a program of an initial
reconnaisance of the plates for early results. He identified over 400 new
peculiar stars (Bidelman 1983, 1985, 1998), but on only a relatively small
subset of the plate collection. Unfortunately, to my knowledge, the remainder of
the northern plates has not been scanned for peculiar stars---let alone all of
the stars given MK classifications.  As Billy remarked in his 1998 paper,
``$\dots$ these plates contain high-quality spectral information on a very large
number of faint stars $\dots$ that should be made available to the astronomical
community. It is to be hoped that plans will be forthcoming from some quarter
for the further utilization of this magnificent observational material.'' The
northern-hemisphere plates are now located at Lowell Observatory (B.~Skiff,
private communication), but I am not aware of a plan to extract their hidden
treasures, which must be numerous. As Carl Sagan said, ``Somewhere, something
incredible is waiting to be known.''

Bidelman formally retired from CWRU in 1985, but continued to come to his office
at the department for many years thereafter (Figure~5). He became interested in
applying astronomical evidence to Biblical events; this included the ``Star of
Bethlehem,'' which he argued was actually a pair of close conjunctions of Venus
and Jupiter (Bidelman 1991). His last publication came in 2009, at the age of
91, and 67 years after his first paper in the {\it Astrophysical Journal}. With
declining health, he moved to Tennessee to be near his daughter Billie, and he
passed away there on 2011 May~3. He had been preceded in death by his wife Verna
and daughter Lana, and is survived by three daughters, eight grandchildren, and
nine great-grandchildren. 

Billy had many interests during his long life---philately, music (he played the
piano well), square-dancing, baseball, plays, movies---but mostly he was a lover
and collector of the stars and their oddities. More than once, after a
conference in his office, I was on the way back up to the floor where the grad
students lived, but he would call me back to take a look at just one more
spectrogram. When I showed him an interesting new object on an objective-prism
plate, he would glance at it but then like a connoisseur he would browse through
the other spectra, and say ``Yes, but what's this one over here?''  He was a
master of the morphological approach to science, content to leave detailed
astrophysical explanations to others. At a conference, he once said  ``I don't
particularly like being referred to as an astrophysicist. I hope that I'm an
astronomer!''

This style of astronomy now recedes into the past. Spectrograms are obtained
with electronic devices and can be made available to everyone in archives,
rather than remaining in analog form on photographic plates in private
collections. Journal papers can be found instantaneously without leaving one's
chair. Massive amounts of information about the stars are stored in machines,
rather than in the brains of astronomers like Bidelman who had mastered the
literature. As I write this, the rueful words of Hamlet, musing on the passing
of his revered father, come to mind: ``I shall not look upon his like again.''

Billy cherished his friendships and collaborations with astronomers from around
the world. He lived in an era of face-to-face conversations and hand-written or
typewritten letters, long before the invention of electronic mail or smartphones
or online preprints. I want to close by recalling some after-dinner remarks at a
conference on the H-R diagram in Washington, DC, in which Bidelman (1977)
reminisced about his long career and friendships: 

{\narrower

``I recall with great pleasure my early contacts with astronomers---at Harvard
$\dots$ and the whole motley Yerkes crew: Struve, Morgan, Greenstein, Henyey,
Chandrasekhar, Kuiper, and all the rest. When I think about these people $\dots$
I marvel at what wonderful people they were---so passionately devoted to
science. But more, they were $\dots$ passionately devoted to life itself.
$\dots$ I've always felt that astronomers on the whole are the best people on
Earth. Let us never forget, nor let our students forget, that of every million
people on the face of the Earth, only one is an astronomer.''

}

%



\begin{figure}
\begin{center}
\includegraphics[height=4.5in]{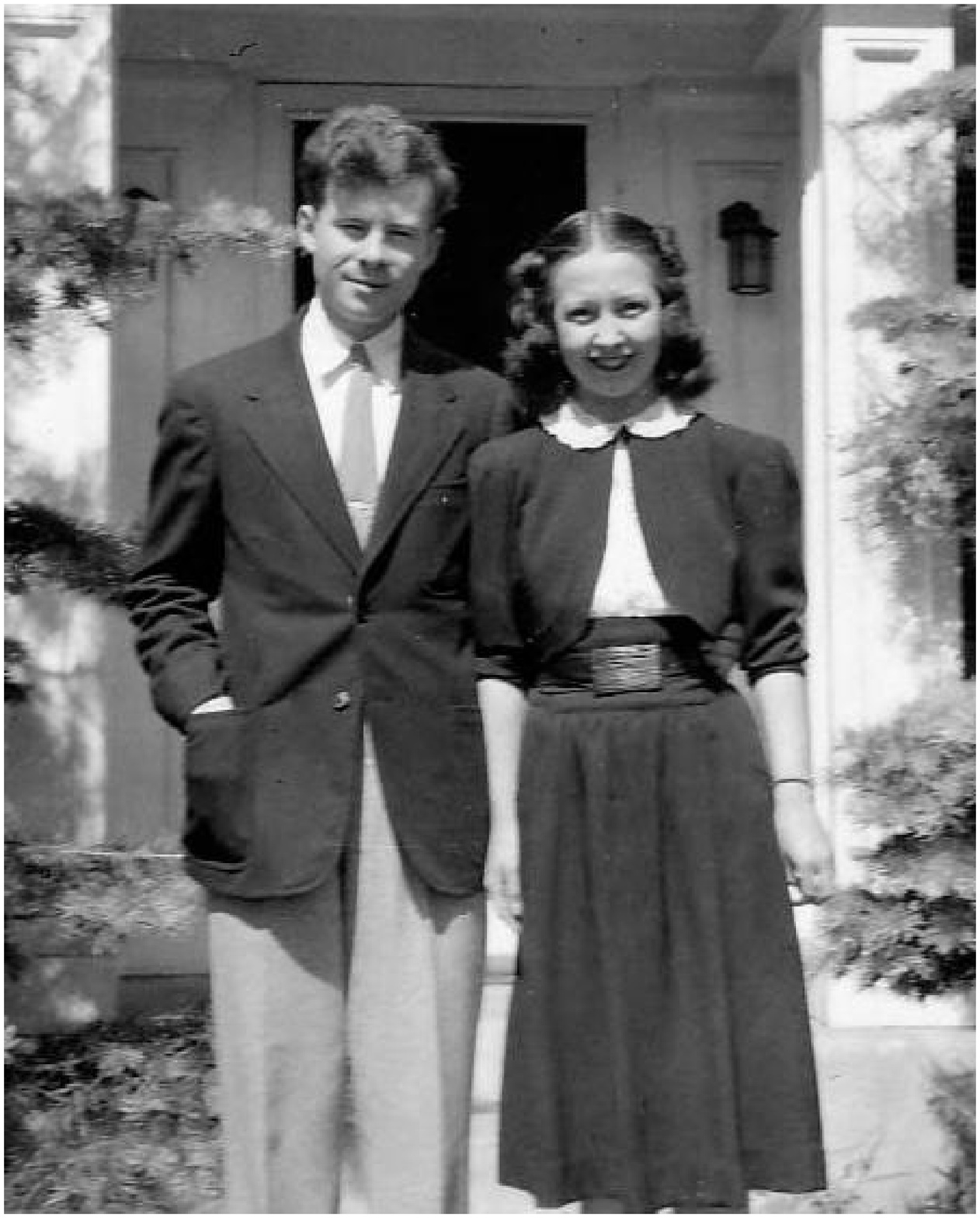}
\figcaption{
William P. Bidelman and Verna S. Bidelman in 1940 June, shortly after their
marriage. {\it Photo courtesy Joseph Little.}
}
\end{center}
\end{figure}

\begin{figure}
\begin{center}
\includegraphics[width=6.5in]{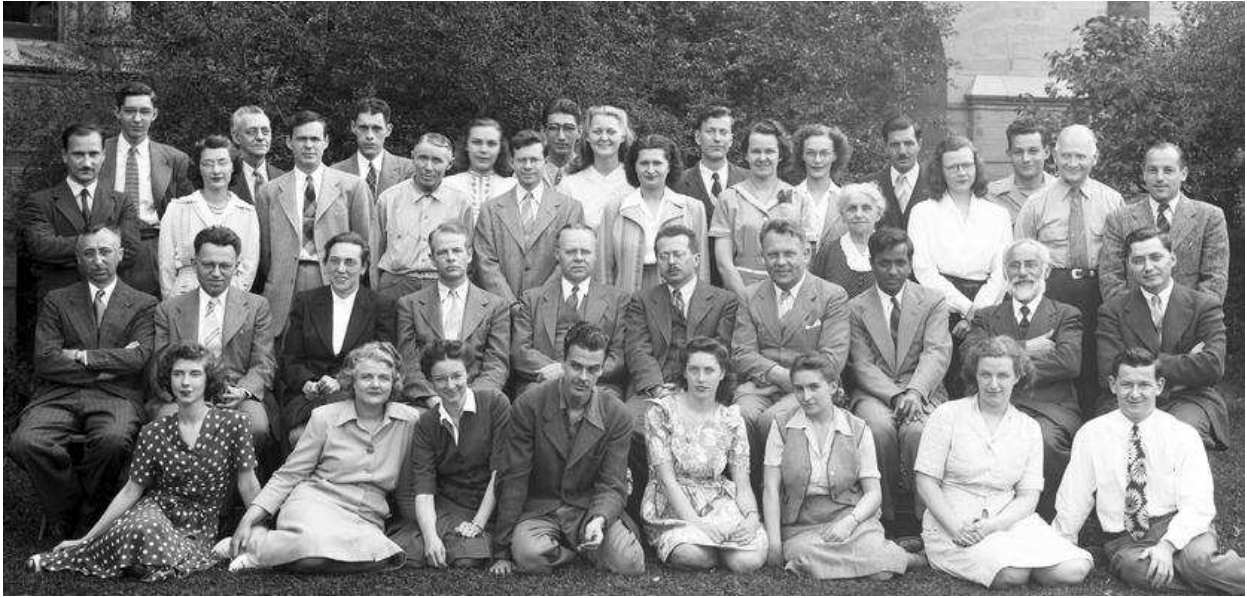}
\figcaption{
The staff of Yerkes Observatory in 1946 May, one the most astonishing groups
of major astronomers of the mid-20th century ever photographed in one place.
Bidelman is fifth from the left in the third row. In the row in front of him can
be seen Swings, Herzberg, Morgan, Struve, Greenstein, Kuiper, Chandrasekhar, and
Van Biesbroeck, among other titans. The front row includes Underhill, M\"unch,
and Roman. In the back row are Slettebak, Deutsch, Wrubel, Code,  Blanco, and
others. {\it University of Chicago Photographic Archive, apf6-00492, Special
Collections Research Center, University of Chicago Library.}
}
\end{center}
\end{figure}

\begin{figure}
\begin{center}
\includegraphics[height=4in]{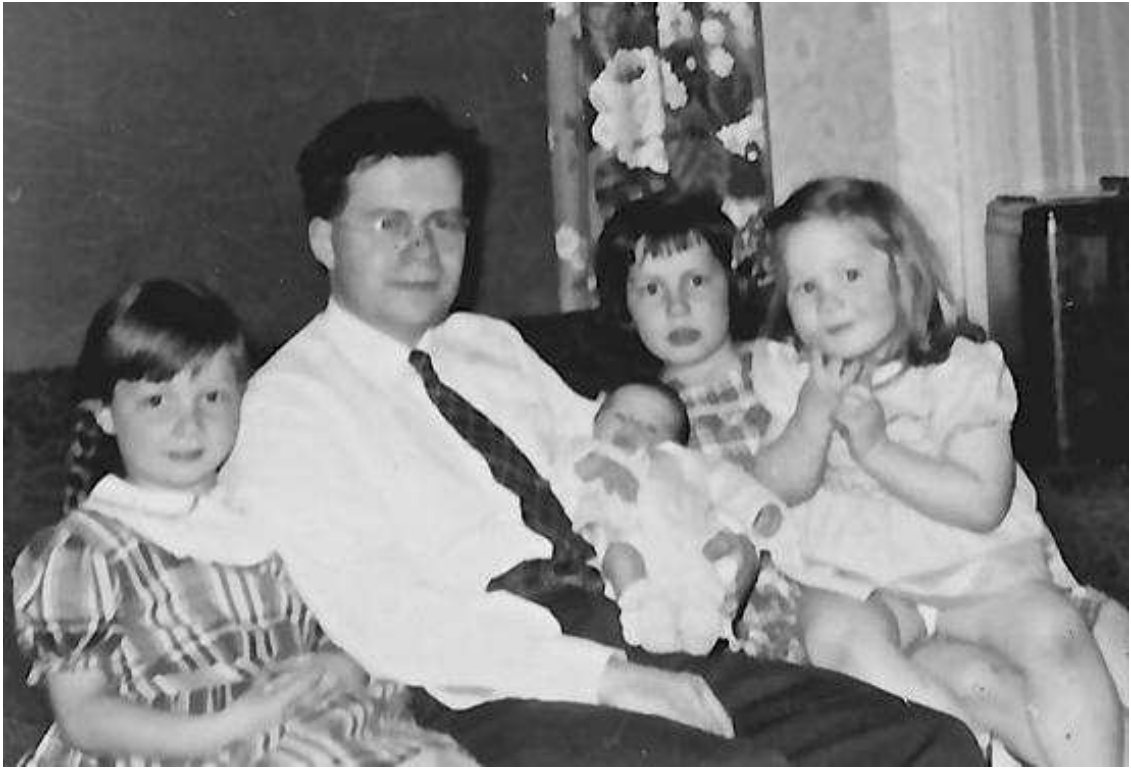}
\figcaption{
Bidelman at his home on Mount Hamilton in 1953 April, shortly after his arrival
at Lick Observatory, with his daughters. Left to right: Linda, Barbara, Lana,
and Billie Jean. {\it Photo courtesy Billie Bidelman Little.}
}
\end{center}
\end{figure}

\begin{figure}
\begin{center}
\includegraphics[height=4in]{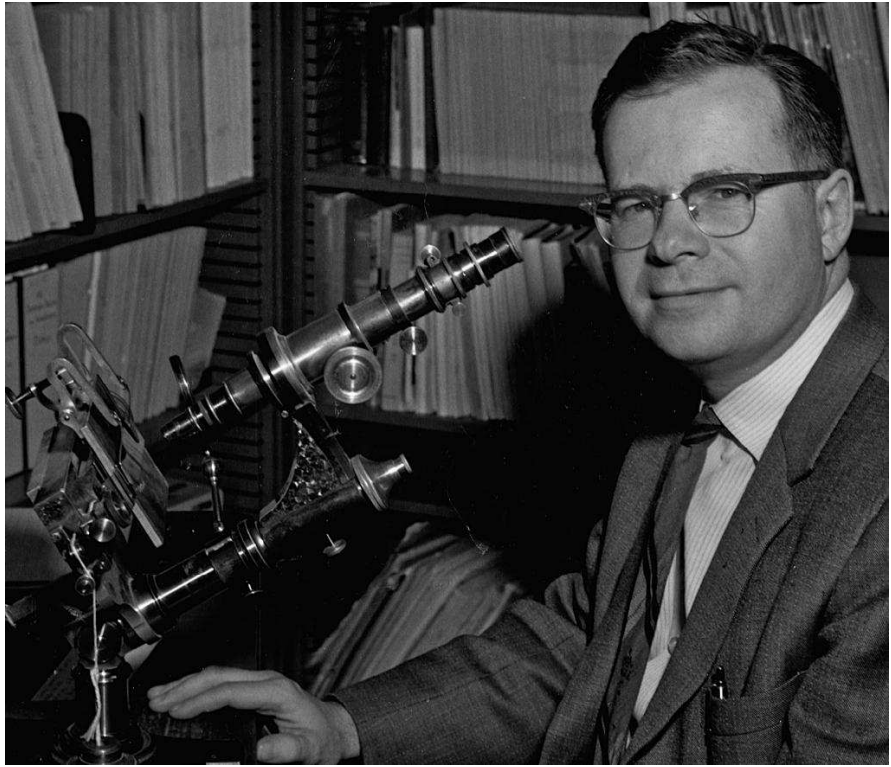}
\figcaption{
A more formal photograph, showing Bidelman examining a spectrogram at Lick. {\it
Photo courtesy Joseph Little.}
}
\end{center}
\end{figure}

\begin{figure}
\begin{center}
\includegraphics[height=4in]{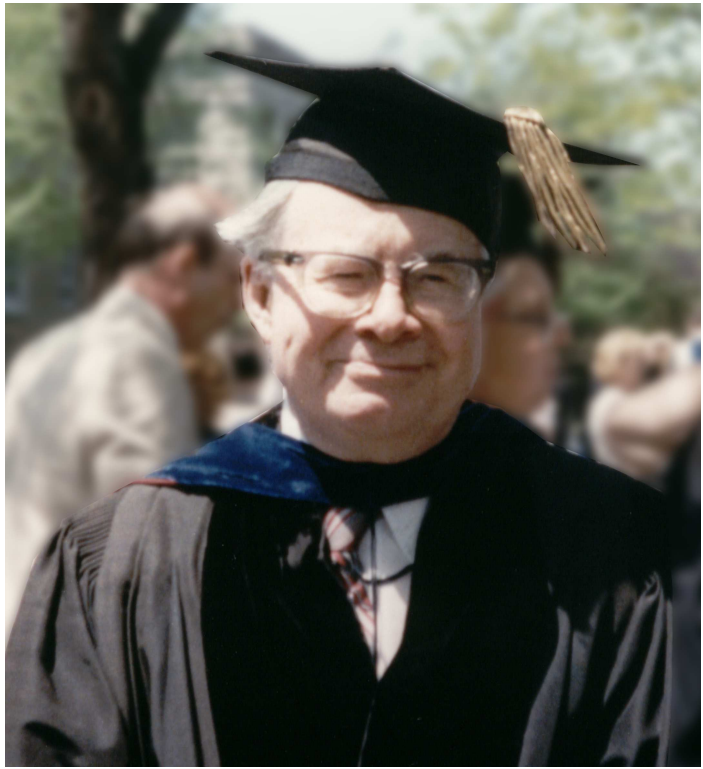}
\figcaption{
Emeritus Professor Bidelman at a Case Western Reserve graduation ceremony in
1988 May, wearing the quizzical expression that many colleagues and students
remember. {\it Photo courtesy James Caplinger.}
}
\end{center}
\end{figure}

\end{document}